\documentclass [aps, preprint, pre,12pt] {article}

\usepackage [utf8] {inputenc}
\usepackage [T1] {fontenc}
\usepackage {geometry}
\usepackage{amsmath}
\usepackage{mathrsfs}
\begin {document}
\title {Susceptibility of the one-dimensional Ising model: is the singularity at T = 0 an essential one?}
\author {James H. Taylor\\
School of Geoscience, Physics, and Safety\\
University of Central Missouri, Warrensburg, MO  64093, U.S.A.\\
jtaylor@ucmo.edu}

\date {May 11, 2022}

\maketitle

\begin {abstract}
The zero-field isothermal susceptibility of the one-dimensional Ising model with nearest-neighbor interactions and a finite number of spins is shown to have a relatively simple singularity as the temperature approaches zero, proportional only to the inverse temperature.  This is in contrast to what is seen throughout the literature for the inifinite chain: an essential singularity that includes an exponential dependence on the inverse temperature. Assuming an arbitrary (but finite) number of spins and retaining terms that are usually considered ignorable in the thermodynamic limit, the analysis involves nothing beyond straightforward series expansions, starting either from the partition function for a closed chain in a magnetic field, obtained using the transfer-matrix approach; or from the expression for the zero-field susceptibility found via the fluctuation-dissipation theorem.  In both cases, the exponential singularity is exactly removed. In addition, the susceptibility per spin is found to increase with the number of spins (except in the case of noninteracting spins), a result which is also at variance with what is normally reported for an infinite chain.
\\
\\
spin chains; Ising model; magnetic susceptibility; finite-size systems; rigorous results in statistical mechanics\\
\\
\\

\end {abstract}

\newpage
\begin{center}
I. INTRODUCTION
\end{center}

The tractability of the one-dimensional (1D) Ising model with only nearest-neighbor interactions \cite{isi} and in zero magnetic field has made it of value not only for pedagogical purposes, but for shedding light on formally corresponding properties of other systems that can be mapped onto it (such as the the folding-unfolding behaviour of proteins \cite{cor}, or the hydrogen-bond order of single-file water in nanopores \cite{kof}). The inclusion of a field complicates the problem mathematically, but allows for even broader applicability: the chemical potential can be substituted for the magnetic field in the treatment of the 1D lattice gas \cite{tok}; a "mean-field" based on experimental measurements of spike trains in chains of neurons can be used to infer parameters describing the interactions between neurons \cite{rou}; and allowing the field to represent a "background"---for example, the existence of dynamical equilibrium between fitness and mutations---has led to uses in population biology, and to an effective alternative to standard approaches in genetic statistical analysis \cite{maj}. Variations have even been applied in fields such as economics \cite{bor} and sociology \cite{mal}.The model has also been employed as a benchmark for testing methods of broader applicability: among more recent examples are a Pad\'e approximant method for finding the thermodynamic properties of a system by first determining the entropy as a function of energy, rather than the free energy as a function of temperature \cite{ber};  approximating the free energy using a cluster-expansion algorithm that discards clusters whose  contributions are below a specified threshold, instead of using a predetermined limit on cluster size \cite{coc}; a combinatorial method that emphasizes the role of nearest-neighbor spin-pairs rather than individual spins \cite{set}; an experimentally realizable technique for finding Lee-Yang zeroes \cite{pen}; and an extended scaling approach designed specifically for spin systems with critical temperature $T_c = 0$ \cite{kat}.

When explicitly referenced, the general concensus is that the isothermal susceptibility for an infinite chain of spins in zero magnetic field has an essential singularity as the absolute temperature $T$ approaches zero, being proportional to the product of the inverse temperature and an exponential whose argument is proportional to $1/T$ as well. In the following, the susceptibility of a chain of finite (but otherewise arbitrary) length is evaluated at low $T$, both for a chain with free ends (open chain) and a chain with periodic boundary conditions (closed chain). Terms in the exact expressions that are usually ignored in the thermodynamic limit are retained. As a result, it is found that the essential singularity as $T \to 0$ is eliminated, leaving a much weaker dependence on the inverse temperature, although at the same time displaying a stronger dependence on the number of spins than is quoted elsewhere.
\\
\\
\begin{center}
II. OPEN CHAIN
\end{center}

As is well-known, the classical 1D Ising model can be described as a chain of spins which can take on two values: $s = +1$ ("up") and $s = -1$ ("down"). For a chain of N spins with uniform, nearest-neighbor-only interactions in zero magnetic field, the Hamiltonian is commonly written in the form 
\begin{equation}
\mathscr{H} = -\sum_{i=1}^{N-1}J(s_is_{i+1}),
\end{equation}
where the interaction strength $J>0$ for a ferromagnetic system. The partition function is then 
\begin{equation}
Z_N = \sum_{\lbrace s_i\rbrace}e^{-\mathscr{H}\beta},
\end{equation}
where $\beta = 1/kT$ ($k$ being Boltzmann's constant), and the sum is over all possible states of the system (i.e, the set of all possible strings of values for the $s_i$). While the transfer-matrix technique may be applied \cite{bei,yqi}, for an open chain the sum can be worked out fairly easily by a combination of direct enumeration and induction (see, for example, \cite{sta} for details):
\begin{equation}
Z_N = 2^N\textrm{cosh}^{(N-1)}(J\beta).
\end{equation}
Once the partition function has been obtained, the spin-spin correlations,
 $\langle s_is_{i+j}\rangle$, can be determined. These may in turn be used to find the zero-field suceptibility via a standard result based on the fluctuation-dissipation theorem \cite{sta}:
\begin{equation}
\chi_T (H = 0,T) = \frac{\mu^2}{kT} \sum_{i = 1}^{N}\sum_{j=1}^N \langle s_is_j\rangle,
\end{equation}
where $H$ is the applied magnetic field and $\mu$ is the effective magnetic moment of a spin. 

By definition,
\begin{equation}
 \langle s_is_{i+l}\rangle = \frac{1}{Z_N}\sum_{\lbrace s_i\rbrace}(s_is_{i+l})e^{-\mathscr{H}\beta},
\end{equation}
and is found in this situation to be equal to $\textrm{tanh}^l(J\beta)$ \cite{sta}. Inserting this result in Eq. (4) yields \cite{sta}, \cite{der}
\begin{equation}
\chi_T (H = 0,T) = \frac{\mu^2}{kT}\bigg\lbrace N \bigg[ 1 + \frac{2\, \textrm{tanh}(J\beta)}{1-\textrm{tanh}(J\beta)}\bigg] - \frac{2 \, \textrm{tanh}(J\beta)[1-\textrm{tanh}^N(J\beta)]}{[1-\textrm{tanh}(J\beta)]^2}\bigg\rbrace.
\end{equation}
Since the last term inside the braces is not proportional to $N$, and since tanh$(J\beta)$ is less than one for finite $J\beta$, that term is normally ignored, on the grounds that in the thermodynamic limit it will be negligible in comparison to the term preceding it. This leads to 
\begin{equation}
\chi_T (H = 0,T) = \bigg( \frac{N\mu^2}{kT} \bigg) e^{2J\beta};
\end{equation}
the susceptibility diverges at low $T$ as $e^{2J\beta}/T$.

Superficially, the same argument seems plausible if $N$ is simply assumed to be extremely large. However, assuming a finite (though otherwise unspecified) value for $N$, this last part can be approximated via a simple series expansion. First, note that
\begin{equation}
\frac{[1 - \textrm{tanh}^N(J\beta)]}{[1 - \textrm{tanh}(J\beta)]} = \sum_{m =0}^{N -1} \textrm{tanh}^m(J\beta);
\end{equation}
also,
\begin{equation}
\textrm{tanh}(J\beta) = \frac{\textrm{sinh}(J\beta)}{\textrm{cosh}(J\beta)} = \frac{(e^{J\beta} - e^{-J\beta})}{(e^{J\beta} + e^{-J\beta})} = \frac{(1 - e^{-2J\beta})}{(1 +e^{-2J\beta})},
\end{equation}
and thus
\begin{equation}
\frac{2 \,\textrm{tanh}(J\beta)}{[1 - \textrm{tanh}(J\beta)]} = e^{2J\beta} - 1.
\end{equation}

Since $e^{-2J\beta}<1$, the factor of $(1 + e^{-2J\beta})^{-1}$ in the last part of Eqs. (9) can be approximated as usual using
\begin{equation}
(1+x)^{-1} = 1-x+x^2-x^3+x^4-...,
\end{equation} 
 leading to 
\begin{multline}
\frac{[1 - \textrm{tanh}^N(J\beta)]}{[1 - \textrm{tanh}(J\beta)]} = \sum_{m =0}^{N -1}[1-e^{-2J\beta}]^m[1 - e^{-2J\beta} + e^{-4J\beta} - e^{-6J\beta} + ...]^m\\ = \sum_{m =0}^{N -1}\bigg[1- 2me^{-2J\beta} + 2m^2e^{-4J\beta} - \frac{2m(2m^2+1)e^{-6J\beta}}{3}\\ + \frac{2m^2(m^2+2)e^{-8J\beta}}{3} - \frac{2m(2m^4+10m^2+3)}{15}e^{-10J\beta} + ...\bigg]
\end{multline}
Using standard results for the sums over different powers of $m$ \cite{gra}, the right-hand side becomes
\begin{multline}
N\bigg[1 - (N-1)e^{-2J\beta} + \frac{(N-1)(2N-1)}{3}e^{-4J\beta} \\ - \frac{(N-1)(N^2 - N +1)}{3}e^{-6J\beta} + \frac{(N - 1)(2N-1)(N^2-N+3)}{15}e^{-8J\beta} \\  - \frac{(N-1)(2N^4-4N^3+16N^2-14N+9)}{45}e^{-10J\beta} + ...\bigg ]
\end{multline}
Utilizing this in Eq. (6) gives
\begin{multline}
\chi_T (H = 0,T) = \frac{N\mu^2}{kT}\bigg[N-(N-1)e^{-2J\beta}-  (N-1)(1-e^{-2J\beta}) \times \\  \times \bigg\lbrace \frac{(2N-1)}{3}e^{-2J\beta} - \frac{(N^2 - N +1)}{3}e^{-4J\beta} + \frac{(2N-1)(N^2-N+3)}{15}e^{-6J\beta}  \\-  \frac{(2N^4-4N^3+16N^2-14N+9)}{45}e^{-8J\beta} + ...\bigg \rbrace \bigg]. 
\end{multline}
After some multiplication and collection of terms, this simplifies to 
\begin{multline}
\chi_T (H = 0,T) = \frac{(N\mu)^2}{kT}\bigg[1-\frac{2(N^2-1)}{3N}e^{-2J\beta} + \frac{(N^2-1)}{3}e^{-4J\beta} \\ -\frac{2(N^2+1)(N^2-1)}{15N}e^{-6J\beta} + \frac{(N^2-1)(2N^2+7)}{45}e^{-8J\beta} - ...\bigg].
\end{multline}
For $Ne^{-2J\beta}<1$, the series inside the square brackets converges; the susceptibility of the chain still diverges as the temperature approaches zero, but only as $1/T$.
Note also that in the special case $J=0$ (no interaction between spins), the factor of $(1-e^{-2J\beta})$ multiplying the term in braces in Eq. (14) becomes zero, and the entire quantity inside the square brackets reduces to $1$, making the zero-field susceptibility per spin just $\mu^2/kT$, as expected. (The same result is obtained for finite $J$ but with $T\to \infty$.) However, for nonzero $J$ and sufficiently low temperature, the susceptibility per spin is (to first order) directly proportional to the number of spins, also in contradiction to what is usually shown. (In a detailed analysis of a finite antiferromagnetic chain with an impurity at one end \cite{yqi}, the authors found that $T\chi_T$ is proportional to the square of the "net spin" of the system; although they provide no details, they mention that they have found the same to be true for other finite chains as well, including a ferromagnetic one with no impurity, which appears to support the result given here.) 
\\
\\
\begin{center}
III. CLOSED CHAIN
\end{center}

It will be assumed initially that the system is in a uniform, nonzero, longitudinal magnetic field $H$. For a chain with periodic boundary conditions ($s_{N+1} \equiv s_{1}$), the transfer-matrix technique leads in a particularly simple fashion to $Z_N$; the partition function thus obtained will be used to find the susceptibility via differentiation with respect to $H$, after which the field will be set to zero.

For the 1 D Ising model with periodic boundary conditions, the transfer-matrix method involves the construction of the matrix $\mathbf{P}$ whose elements have the form
\begin{equation}
\textrm{exp}[-U(s_i,s_{i+1})\beta],
\end{equation}
where
\begin{equation}
U(s_i,s_{i+1}) = -Js_is_{i+1} - \frac{\mu H}{2}(s_i+s_{i+1}),
\end{equation}
with $s_i$ and $s_{i+1}$ taking on the values $\pm1$:
\\
\begin{equation}
\mathbf{P} = \left[
\begin{array}{ccc}
e^{(J + \mu H)\beta} & e^{-J\beta} \\
e^{-J\beta} & e^{(J - \mu H)\beta}
\end{array} \right]
\end{equation}
\\
\cite{bei, sta, rob}. Because of the boundary conditions, the partition function can be expressed as the trace of the product of $N$ such matrices. The result is simply the sum of the eigenvalues of $\mathbf{P}$, each raised to the power $N$. The eigenvalues are easily found to be
\begin{equation}
\lambda_\pm = e^{J\beta}\textrm{cosh}(\mu H\beta) \pm  \sqrt{e^{2J\beta}\textrm{cosh}^2(\mu H\beta) -2 \, \textrm{sinh}(2J\beta)},
\end{equation}
so
\begin{equation}
Z_N = \lambda_+^N + \lambda_-^N = \lambda_+^N \bigg[ 1 +\bigg( \frac{\lambda_-}{\lambda_+} \bigg)^N \bigg].
\end{equation}
At this point it is usually assumed that, in the thermodynamic limit, the second term can be ignored in comparison to the first, since the ratio $(\lambda_-/\lambda_+)$ is always less than 1. However, if $H=0$, this assumption is no longer true at $T = 0$, and is of questionable appropriateness for very small $T$, as shown below.

The susceptibility is given by  
\begin{equation}
\chi_T (H,T) = \bigg( \frac{\partial M}{\partial H}\bigg)_T = \frac{\partial}{\partial H}\bigg[\frac{\partial [kT\mathrm{ln}(Z_N)]}{\partial H}\bigg]_T,
\end{equation}
where $M$ is the magnetization of the chain. Starting from Eq. (20) (including both terms), and setting $H = 0$ after performing the required differentiations,  one obtains
\begin{equation}
\chi_T(H = 0,T) = \bigg(\frac{N\mu^2}{kT}\bigg)e^{2J\beta}\bigg[\frac{\textrm{cosh}^N(J\beta) - \textrm{sinh}^N(J\beta)}{\textrm{cosh}^N(J\beta) + \textrm{sinh}^N(J\beta)} \bigg].
\end{equation}
Assuming that $\textrm{sinh}^N(J\beta)$ can be ignored in comparison to $\textrm{cosh}^N(J\beta)$ again leads to Eq. (7) for the zero-field susceptibility.   However, it may be noted that
\begin{multline}
\frac{\textrm{cosh}^N(J\beta) - \textrm{sinh}^N(J\beta)}{\textrm{cosh}^N(J\beta) + \textrm{sinh}^N(J\beta)} = \frac{(e^{J\beta} + e^{-J\beta})^N -  (e^{J\beta} - e^{-J\beta})^N}{(e^{J\beta} + e^{-J\beta})^N + (e^{J\beta} - e^{-J\beta})^N}\\ = \frac{(1 + e^{-2J\beta})^N -  (1 - e^{-2J\beta})^N}{(1 + e^{-2J\beta})^N + (1 - e^{-2J\beta})^N}.
\end{multline}
Expanding the factors of $(1 \pm e^{-2J\beta})^N$ yields
\begin{multline}
\frac{\textrm{cosh}^N(J\beta) - \textrm{sinh}^N(J\beta)}{\textrm{cosh}^N(J\beta) + \textrm{sinh}^N(J\beta)} =\\ Ne^{-2J\beta}\Bigg[1 + \sum_{\substack{m =3 \\ m \ odd}}^N \frac{(N-1)!}{m!(N-m)!}e^{-2(m-1)J\beta}\Bigg] \Bigg[1 + \sum_{\substack{n = 2 \\n \  even}}^N \frac{N!}{n!(N-n)!}e^{-2nJ\beta}\Bigg]^{-1}. 
\end{multline}
Inserting this in Eq. (22) gives 
\begin{multline}
\chi_T(H = 0,T) = \\ \frac{(N\mu)^2}{kT}\Bigg[1 + \sum_{\substack{m =3 \\ m \ odd}}^N \frac{(N-1)!}{m!(N-m)!}e^{-2(m-1)J\beta}\Bigg] \Bigg[1 + \sum_{\substack{n = 2 \\n \  even}}^N \frac{N!}{n!(N-n)!}e^{-2nJ\beta}\Bigg]^{-1}
\end{multline}
This expression is exact for any finite $N$. As for the open chain, the essential singularity has been removed, and the susceptibility per spin clearly increases with $N$ when $Ne^{-2J\beta}$ is sufficiently small.

Nevertheless, it may be useful to develop a series approximation here, similar to that given in Eq. (15). Note that the denominator is again of the form $(1+x)$, so if x (here, the sum over $n$ in the denominator) is less than 1, the approximation in Eq. (11) will converge. To determine whether or not this is the case, note that
\begin{equation}
\sum_{n=0}^{\infty}\frac{1}{n!} = e,
\end{equation}
and that
\begin{multline}
\frac{N!}{n!(N-n)!}e^{-2nJ\beta} = \frac{N(N-1)...(N-n+1)}{n!}e^{-2nJ\beta} \\ = \frac{[Ne^{-2J\beta}][(N-1)e^{-2J\beta}]...[(N-n+1)e^{-2J\beta}]}{n!} < \frac{[Ne^{-2J\beta}]^n}{n!}
\end{multline}
for $n>1$. Assuming $Ne^{-2J\beta}=1$,
\begin{equation}
 \sum_{\substack{n = 2 \\n \  even}}^N \frac{N!}{n!(N-n)!}e^{-2nJ\beta} < \sum_{\substack{n = 2 \\n \  even}}^N \frac{1}{n!} < \sum_{n=2}^{\infty}\frac{1}{n!} = e-2 \simeq.718282.
\end{equation}
Clearly, then, if $Ne^{-2J\beta}<1$, Eq. (11) can be used to approximate the last of the three terms on the right-hand side of  Eq. (24). Multiplying through, one obtains
\begin{equation}
Ne^{-2J\beta}\bigg[1-\frac{(N^2-1)}{3}e^{-4J\beta} + \frac{(2N^2-3)(N^2-1)}{15}e^{-8J\beta} -...\bigg].
\end{equation}

Finally, inserting this expression in Eq. (21) yields.
\begin{equation}
\chi_T (H = 0,T) = \frac{(N\mu)^2}{kT}\bigg\lbrace1 - \frac{(N^2 - 1)}{3}e^{-4J\beta} + \frac{(2N^2-3)(N^2-1)}{15}e^{-8J\beta} -...\bigg\rbrace. 
\end{equation}
Note that for the closed chain, the susceptibility is a power series in $e^{-4J\beta}$, whereas in the case of the open chain the result includes \textit{all} powers of $e^{-2J\beta}$, not only the even ones. This difference, of course, is a direct result of the difference in boundary conditions.

In the noninteracting case, Eq. (30) is not particularly useful since the series expansion in Eq. (11) does not converge. However, referring back to Eq. (24), it can be seen that for $J=0$ the numerator and denominator are both sums over binomial coefficients:
\begin{equation}
N\Bigg[1 + \sum_{\substack{m =3 \\ m \ odd}}^N \frac{(N-1)!}{m!(N-m)!}\Bigg] = \sum_{\substack{m =1 \\ m \ odd}}^N \frac{(N)!}{m!(N-m)!}
\end{equation}
and
\begin{equation}
1 + \sum_{\substack{n =2 \\n \  even}}^N \frac{N!}{n!(N-n)!} =  \sum_{\substack{n =0 \\n \  even}}^N \frac{N!}{n!(N-n)!}.
\end{equation}
Both sums are equal to $2^{(N-1)}$ \cite{gra}, so the expression on the right-hand side of Eq. (24) becomes equal to $1$, and the susceptibility per spin is again $\mu^2/kT$. As for the open chain, the same result is obviously obtained for finite $J$ as $T \to \infty$. 
\\
\\
\begin{center}
IV. CONCLUSIONS
\end{center}

For finite-length chains of Ising spins---both open and closed---the singularity in the zero-field susceptibility as $T \to 0$ appears not to have the form that has been generally accepted for the infinite chain. Instead of diverging as $\beta e^{2J\beta}$, $\chi_T  \sim \beta$ (to leading order) when $Ne^{-2J\beta}$ is small. Also, the susceptibility per spin is found to be directly proportional to the number of spins (also to leading order), rather than being independent of the chain length.

The final expressions given for the susceptibility in sections II and III remain useful as long as $Ne^{-2J\beta} < 1$. Obviously, what this requirement amounts to is a restriction on the range of temperatures for which this is correct, given the value of $N$ (i.e., $T<2J/k\mathrm{ln}N)$. But this limitation is not really very severe. For example, if $N=10^{10}$, it might seem quite reasonable to ignore the term $(\lambda_-/\lambda_+)^N$ in Eq. (20), even assuming that $(\lambda_-/\lambda_+)$ is quite close to unity; however, letting $J = .01$ eV \cite{kit} (a rough typical value for the interaction between spins in real ferromagnets), the first few terms displayed explicitly in Eq. (30) would still be a fair approximation for any $T$ below $10$ K. Increasing the number of spins to $10^{1000}$ only causes the upper limit to drop by a factor of $100$, to $0.1$ K.  Even though the allowable range of temperature ultimately becomes infinitesimal as $N$ continues to increase, it obviously always includes $T = 0$.

It should be noted that for higher temperatures ($T\gg J/k$), sinh$(J\beta)$ will be very small compared to cosh$(J\beta)$. In that case, the last term in Eq. (6) is indeed ignorable, and the result for the susceptibility that is generally quoted for the infinite chain---Eq. (7)---should be a good approximation. Simlarly, for large values of $N$, sinh$^N(J\beta)$ will be insignificant in comparison to cosh$^N(J\beta)$ in Eq. (22), leading to the same result.

It can be seen from the preceding that simple and seemingly reasonable arguments about the effects of going to the thermodynamic limit are not reinforced in any obvious way by the finite chain results. There does not appear to be a clear reason why the apparent form of the susceptibility at low temperatures should change so dramatically in the limit as the number of spins "goes to infinity."

\end {document}